\documentclass[useAMS,usenatbib]{mn2e}

\usepackage{natbib}
\usepackage{graphicx}
\usepackage{amssymb}
\usepackage{bm}
\usepackage{color}
\usepackage[dvipsnames]{xcolor}
\usepackage{caption}
\usepackage{lipsum,graphicx,multicol}
\usepackage{float}
\usepackage[fleqn]{amsmath}
\usepackage{subcaption}
\usepackage{array}

\title[X-ray weak JWST-AGN]{Another view into JWST-discovered X-ray weak AGNs via radiative dusty feedback} 

\author[ ]
{W. Ishibashi$^{1,2,3}$\thanks{E-mail: wako.ishibashi@physik.uzh.ch}, A. C. Fabian$^{4}$, R. Maiolino$^{5}$, Y. Gursahani$^{6}$ and C. S. Reynolds$^{6}$ 
\vspace{0.25cm}
\footnotemark[0]\\
$^{1}$Physik-Institut, Universit$\ddot{a}$t Z$\ddot{u}$rich, Winterthurerstrasse 190, 8057 Z$\ddot{u}$rich, Switzerland \\
$^{2}$Istituto Ricerche Solari (IRSOL), Universit$\grave{a}$ della Svizzera italiana (USI), 6605 Locarno Monti, Switzerland \\
$^{3}$Euler Institute, Universit$\grave{a}$ della Svizzera italiana (USI), 6900 Lugano, Switzerland \\ 
$^{4}$Institute of Astronomy, Madingley Road, Cambridge CB3 0HA \\
$^{5}$Kavli Institute for Cosmology, University of Cambridge, Madingley Road, Cambridge, CB3 OHA \\
$^{6}$Department of Astronomy, University of Maryland, College Park, MD 20742-2421, USA
}

\begin{document}

\pdfminorversion=4

\date{Accepted 2025 August 26. Received 2025 August 7; in original form 2025 May 25} 

\pagerange{\pageref{firstpage}--\pageref{lastpage}} \pubyear{2012}

\maketitle

\label{firstpage}

\begin{abstract}
JWST has revealed a previously unknown population of low-luminosity active galactic nuclei (AGN) in the early Universe. These JWST-AGN at high redshifts are characterised by a set of peculiar properties, including unusually weak X-ray emission. Here we investigate the apparent lack of X-ray emission in the framework of the ``AGN radiative dusty feedback'' scenario based on the effective Eddington limit for dust. We analyse how the boundary in the $N_\mathrm{H} - \lambda$ plane, defined by the column density versus the Eddington ratio, is modified as a function of the dusty gas parameters (metallicity, dust grain size and composition). Low metallicity gas with little dust content tends to survive against radiation pressure, and likely accumulates in the nuclear region. We suggest that such dust-poor gas can provide long-lived absorption and may lead to heavy X-ray obscuration, as observed in early JWST-AGN. The blowout vs. stalling condition of the obscuring clouds indicates that higher metallicities are required to eject heavier column densities, while large columns of gas can stall in low metallicity environments. Therefore the metallicity may play a key role in the AGN radiative dusty feedback scenario. We discuss how other peculiar properties of JWST-AGN --such as Balmer absorption features and weak radio emission-- may be naturally interpreted within the same physical framework. 
\end{abstract}   



\section{Introduction}
\label{Sect_introduction}

Recent JWST observations have uncovered a population of low-to-moderate luminosity ($L \sim 10^{44} - 10^{46}$ erg/s) active galactic nuclei (AGN) at very high redshifts ($z \sim 4 - 11$), hereafter JWST-AGN \citep[e.g.][]{Harikane_et_2023, Greene_et_2024, Matthee_et_2024, Kocevski_et_2023, Kocevski_et_2024, Juodzbalis_et_2024, Maiolino_et_2024, Maiolino_et_2025}. Compared to the high-luminosity massive quasars discovered in the pre-JWST era \citep[][and references therein]{Inayoshi_et_2020, Fan_et_2023}, these moderate AGNs are likely powered by lower mass black holes ($M_\mathrm{BH} \sim 10^6 - 10^8 M_{\odot}$), and are more representative of the underlying  black hole population in the early Universe. Indeed, their number densities are quite high ($\gtrsim 10^{-5} \, \mathrm{Mpc^{-3} mag^{-1}}$) and may account for a few percent of the galaxy population at similar redshifts \citep{Matthee_et_2024, Greene_et_2024, Kocevski_et_2024}. 

A major fraction of JWST-AGN are broad-line type-1 AGN identified by the detection of permitted broad $\mathrm{H \alpha}$ and/or $\mathrm{H \beta}$ emission lines (with FWHM $\gtrsim 1000$ km/s), without a broad counterpart in the forbidden [OIII] line \citep[e.g.][]{Maiolino_et_2025}. Thus the broad component should arise from the broad line region (BLR) rather than from an outflow origin. 

A subset of the newly-discovered JWST-AGN are known as Little Red Dots (LRDs), which form a fraction of around $\sim (10 - 30) \%$ of the type-1 AGN population \citep{Matthee_et_2024, Kocevski_et_2024, Hainline_et_2025}. LRDs are red compact sources characterised by a V-shaped spectral energy distribution, with a red continuum in the rest-frame optical and a blue excess in the rest-frame ultraviolet \citep[][]{Kocevski_et_2023, Labbe_et_2025}. The LRDs could be interpreted as dust-reddened AGNs, and their unique characteristics have attracted much attention in the recent literature. 

A number of peculiar properties are observed in LRDs and more in general the JWST-AGN population at high redshifts: lack of X-ray detection \citep{Ananna_et_2024, Yue_et_2024, Maiolino_et_2025}, weak radio emission \citep{Mazzolari_et_2024b}, and Balmer absorption features \citep{Matthee_et_2024, Kocevski_et_2024, Juodzbalis_et_2024}. In particular, the X-ray weakness is a puzzling result, given that most JWST-AGN are undetected in X-ray observations, even in stacking analysis \citep[][and references therein]{Maiolino_et_2025}.

Two theoretical scenarios have been proposed to explain the apparent lack of X-ray emission from early JWST-AGN: intrinsically weak X-ray emission due to super-Eddington accretion \citep[][see also \citet{Maiolino_et_2025}]{Pacucci_Narayan_2024, Madau_Haardt_2024, King_2025}; and heavy X-ray absorption in dense dust-free gas, such as BLR clouds along the line of sight \citep{Maiolino_et_2025, Juodzbalis_et_2024}. 

Here we investigate the X-ray weakness problem of JWST-AGN in the framework of the ``radiative dusty feedback'' scenario based on radiation pressure on dust. We explore what physical insight may be gained by the analysis of the $N_\mathrm{H} - \lambda$ plane \citep{Fabian_et_2008, Fabian_et_2009, Ishibashi_et_2018b} as a function of the dusty gas properties, and we examine the blowout vs. stalling condition of obscuring clouds. 
The paper is structured as follows. We first recall the basics of the effective Eddington limit for dust, and conduct a thorough analysis of the $N_\mathrm{H} - \lambda$ plane and its dependence on the dusty gas parameters (Section \ref{Sect_NH_lambda}). We then consider the X-ray weak JWST-AGN in the $N_\mathrm{H} - \lambda$ plane perspective (Section \ref{Sect_JWST_AGN}). The blowout vs. stalling conditions of obscuring clouds are analysed in Section \ref{Sect_Clouds}. We discuss additional peculiar features of JWST-AGN, and compare them with other AGN populations in both the local and the early Universe (Section \ref{Sect_Discussion}). 


\section{The $N_\mathrm{H} - \lambda$ plane: dependence on dusty gas properties}
\label{Sect_NH_lambda}


\subsection{Effective Eddington limit in the $N_\mathrm{H} - \lambda$ plane}

We consider AGN radiation pressure acting on the surrounding dusty gas in the gravitational potential of the central black hole. We assume cold neutral dusty gas, where electron scattering is negligible compared to dust absorption. The inward gravitational force on the dusty gas shell is given by
\begin{equation}
F_\mathrm{grav} = 4 \pi G m_p M_\mathrm{BH} N ,
\end{equation}
where $G$ is the gravitational constant, $m_p$ is the proton mass, $N$ is the column density of the dusty gas shell, and $M_\mathrm{BH}$ is the central black hole mass, which is assumed to dominate the local gravitational potential. 
The outward radiative force acting on the same column of dusty gas is 
\begin{equation}
F_\mathrm{rad} = \frac{L}{c} \left(  \tau_\mathrm{IR} + 1 - e^{-\tau_\mathrm{UV}} \right) , 
\end{equation}
where $L$ is the central luminosity, $\tau_\mathrm{IR} = \kappa_\mathrm{IR} m_p N$ and $\tau_\mathrm{UV} = \kappa_\mathrm{UV} m_p N$ are the infrared (IR) and ultraviolet (UV) optical depths, with $\kappa_\mathrm{IR}$ and $\kappa_\mathrm{UV}$ being the dust opacities in the IR and UV bands, respectively. By balancing the radiative and gravitational forces, we derive the effective Eddington luminosity for dusty gas, $L_\mathrm{E}^{'}$. The corresponding effective Eddington ratio for dusty gas is given by
\begin{equation}
\Lambda = \frac{L}{L_\mathrm{E}^{'}} = \frac{L( \tau_\mathrm{IR} + 1 - e^{-\tau_\mathrm{UV}})}{4 \pi G c m_p M_\mathrm{BH} N} . 
\end{equation}

As discussed in our previous papers \citep{Ishibashi_Fabian_2016b, Ishibashi_et_2018a}, three distinct physical regimes can be identified according to the optical depth of the medium: optically thick to both IR and UV, optically thick to UV but optically thin to IR (single scattering limit), and optically thin to both UV and IR. The effective Eddington ratios in the three optical depth regimes are respectively given by
\begin{equation}
\Lambda_\mathrm{IR} = \frac{\kappa_\mathrm{IR} L}{4 \pi G c M_\mathrm{BH} } \quad \mathrm{(IR \, regime)}
\end{equation}
\begin{equation}
\Lambda_\mathrm{SS} = \frac{ L}{4 \pi G c m_p M_\mathrm{BH} N } \quad \mathrm{(SS \, limit)}
\end{equation}
\begin{equation}
\Lambda_\mathrm{UV} = \frac{\kappa_\mathrm{UV} L}{4 \pi G c M_\mathrm{BH} }  \quad \mathrm{(UV \, regime)}
\end{equation}
We note that in the single scattering limit, the effective Eddington ratio is independent of the dust opacities, while it is inversely proportional to the column density of the dusty gas ($\Lambda_\mathrm{SS} \propto 1/N$). In contrast, the effective Eddington ratios are independent of the gas column density in both the IR and UV regimes, while they directly scale with the IR and UV opacities in the IR and UV domains respectively ($\Lambda_\mathrm{IR,UV} \propto \kappa_\mathrm{IR,UV}$). 

Dusty gas is ejected when the central AGN luminosity exceeds the effective Eddington limit. 
This is set by the physical condition $\Lambda = 1$, which translates into a condition on the dusty gas column density
\begin{equation}
N_\mathrm{E} = \frac{( \tau_\mathrm{IR} + 1 - e^{-\tau_\mathrm{UV}})}{\sigma_T} \lambda , 
\label{Eq_NE_lambda}
\end{equation}
where we have introduced the standard Eddington luminosity $L_\mathrm{E} = (4 \pi G c m_p M_\mathrm{BH}) / \sigma_T$ and the classical Eddington ratio $\lambda = L/L_\mathrm{E}$. 

The effective Eddington limit for dusty gas defines a critical boundary in the $N_\mathrm{H} - \lambda$ plane separating the region of long-lived obscuration (to the left of the boundary curve) and the so-called forbidden region (to the right of the boundary curve). The actual location of the boundary in the $N_\mathrm{H} - \lambda$ plane is determined by the dust opacities in the UV and IR bands. 
The UV and IR dust opacities are given by \citep{Thompson_Heckman_2024}
\begin{equation}
\kappa_\mathrm{UV} = \frac{3}{4} \frac{f_\mathrm{dg}}{\rho_d} \frac{1}{\left( a_\mathrm{min} a_\mathrm{max} \right)^{1/2}}
\label{Eq_kappa_UV}
\end{equation}
\begin{equation}
\kappa_\mathrm{IR} = 0.0125 f_\mathrm{dg} T_r^2 , 
\label{Eq_kappa_IR}
\end{equation} 
where $f_\mathrm{dg}$ is the dust-to-gas ratio, $\rho_d$ is the dust grain density, $(a_\mathrm{min}, a_\mathrm{max})$ are the minimum and maximum sizes assuming the Mathis, Rumpl $\&$ Nordsieck (MRN) grain size distribution \citep{Mathis_et_1977}, and $T_r$ is the radiation temperature.

The standard MRN grain size distribution follows a power-law relation of the form $dn/da \propto a^{-3.5}$, with $a_\mathrm{min} = 0.005 \, \mathrm{\mu m}$ and $a_\mathrm{max} = 0.25 \, \mathrm{\mu m}$. A typical grain density may be $\rho_d = 3 \, \mathrm{g \, cm^{-3}}$, an intermediate value between graphite ($\rho_d = 2.26 \, \mathrm{g \, cm^{-3}}$) and silicate ($\rho_d = 3.3 \, \mathrm{g \, cm^{-3}}$) grains. In the IR regime, the optical depth can be approximated by the temperature-dependent Rosseland mean opacity, with a characteristic temperature of $T_r = 200$ K. 

With respect to the dust content of the gas, observations indicate a power-law relationship between the dust-to-gas ratio ($f_\mathrm{dg}$) and the metallicity\footnote{The total metallicity in units of the solar metallicity is given by $Z \cong 2.04 \cdot 10^{-9} \times 10^{12 + \log\mathrm{(O/H)}} Z_{\odot}$, with $12 + \log\mathrm{(O/H)}_{\odot} = 8.69$ and $Z_{\odot} = 0.0134$, assuming solar elemental abundance \citep{Asplund_et_2009, Galliano_et_2021}.} ($Z$) , which may be parametrised by a single power-law of the form
\begin{equation}
\log f_\mathrm{dg} = 1.30 \times [12 + \log(\mathrm{O/H})] - 13.72 
\label{Eq_fdg_Z}
\end{equation}
\citep{DeVis_et_2019, Popping_Peroux_2022}. 
The linear relation between dust-to-gas ratio and gas phase metallicity --traced by oxygen abundance-- seems to hold over cosmic time, with little evolution from $z \sim 0$ to $z \sim 5$ \citep{Popping_Peroux_2022}.  
Using the empirical $f_{dg} - Z$ relation (equation \ref{Eq_fdg_Z}), we can compute the dust-to-gas ratios corresponding to different metallicities. This directly implies a metallicity dependence for the effective Eddington ratios in the IR and UV regimes (since $\Lambda_\mathrm{IR,UV} \propto \kappa_\mathrm{IR,UV} \propto f_\mathrm{dg} \propto Z$).  


\subsection{Variations of the dusty gas parameters}

We recall that the critical curve in the $N_\mathrm{H} - \lambda$ plane is defined by the condition $N = \lambda ( \tau_\mathrm{IR} + 1 - e^{-\tau_\mathrm{UV}} ) / \sigma_T$ (equation \ref{Eq_NE_lambda}), where $\tau_\mathrm{UV} = \kappa_\mathrm{UV} m_p N$ and $\tau_\mathrm{IR} = \kappa_\mathrm{IR} m_p N$, with the UV and IR opacities given in equations \ref{Eq_kappa_UV} and \ref{Eq_kappa_IR}, respectively. The gas column density is assumed to be mostly neutral, also consistent with the emerging picture of the central AGN being embedded in dense neutral gas \citep{Inayoshi_Maiolino_2025}.

We now analyse how variations in the dusty gas properties ($Z$, $a_\mathrm{min}$, $a_\mathrm{max}$, $\rho_d$, $T_r$) affect the boundaries in the $N_\mathrm{H} - \lambda$ plane. 
In the following, we assume as fiducial values $Z = 1 Z_{\odot}$, $\rho_d = 3 \, \mathrm{g \, cm^{-3}}$, $a_\mathrm{min} = 0.005 \mathrm{\mu m}$, $a_\mathrm{max} = 0.25 \mathrm{\mu m}$, $T_r = 200$ K, and vary each physical parameter in turn. 


\subsubsection{Variations in metallicity ($Z$)}

\begin{figure}
\begin{center}
\includegraphics[angle=0,width=0.4\textwidth]{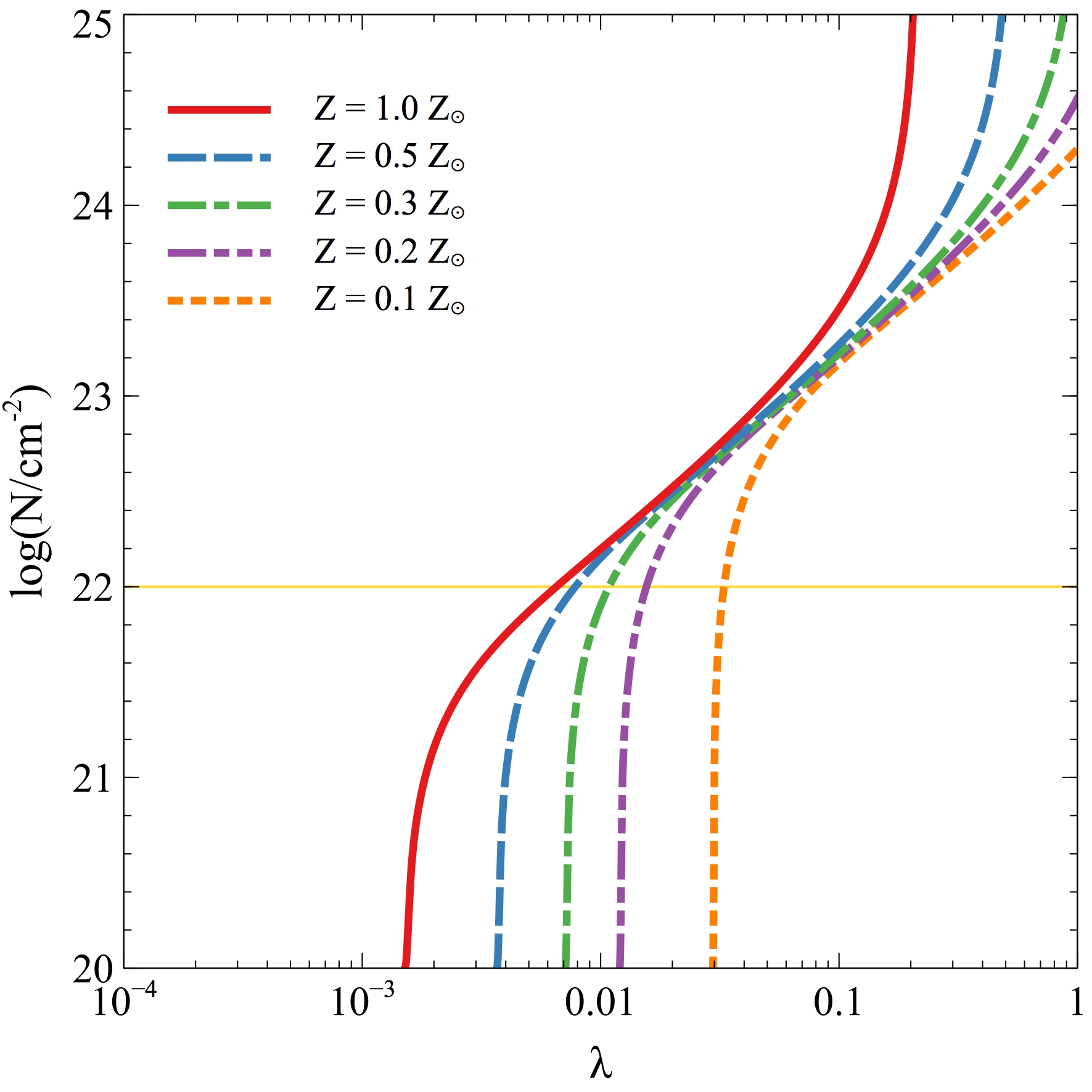} 
\caption{ $N_\mathrm{H} - \lambda$ plane for fiducial parameters and variations in metallicity: $Z = 1 Z_{\odot}$ (red solid), $Z = 0.5 Z_{\odot}$ (blue dashed), $Z = 0.3 Z_{\odot}$ (green dash-dot), $Z = 0.2 Z_{\odot}$ (violet dash-dot-dot), $Z = 0.1 Z_{\odot}$ (orange dotted). The horizontal line (yellow fine line) marks the $N = 10^{22} \mathrm{cm^{-2}}$ limit, below which absorption by outer dust lanes becomes important \citep{Fabian_et_2008, Fabian_et_2009}. 
}
\label{Fig_NH_lambda_var-Z}
\end{center}
\end{figure}

In Fig. \ref{Fig_NH_lambda_var-Z}, we show the $N_\mathrm{H} - \lambda$ plane for different metallicities, from solar to sub-solar values (from $Z = 1 Z_{\odot}$ to $Z = 0.1 Z_{\odot}$). Differences in metallicity give rise to different dust-to-gas ratios (through the linear $f_\mathrm{dg} - Z$ relation), which then determine the effective Eddington ratio. We see that variations in metallicity can lead to significant changes in the $N_\mathrm{H} - \lambda$ plane: lower metallicities move the boundary curve to the right. This reduces the area of the forbidden region, and simultaneously expands the long-lived obscuration region. Both the UV and IR regimes are affected as $f_\mathrm{dg}$ (or equivalently $Z$) appears in both regimes. In contrast, the boundary curves at different metallicities tend to overlap at intermediate column densities, since $\Lambda_\mathrm{SS}$ is independent of $f_\mathrm{dg}$ (or $Z$) in the single scattering limit. 

With decreasing metallicity, the area of the forbidden region decreases, and at the same time the region of long-lived obscuration increases. At lower metallicities, higher Eddington ratios ($\lambda$) would be required to eject a given column density ($N$). This means that radiation pressure cannot easily clear out metal-poor or dust-poor gas, which tends to survive against radiative feedback. The resulting long-lived clouds may stall in the nuclear region and provide prolonged gas absorption. 

\begin{figure*}
\begin{multicols}{2}
\includegraphics[width=0.8\linewidth]{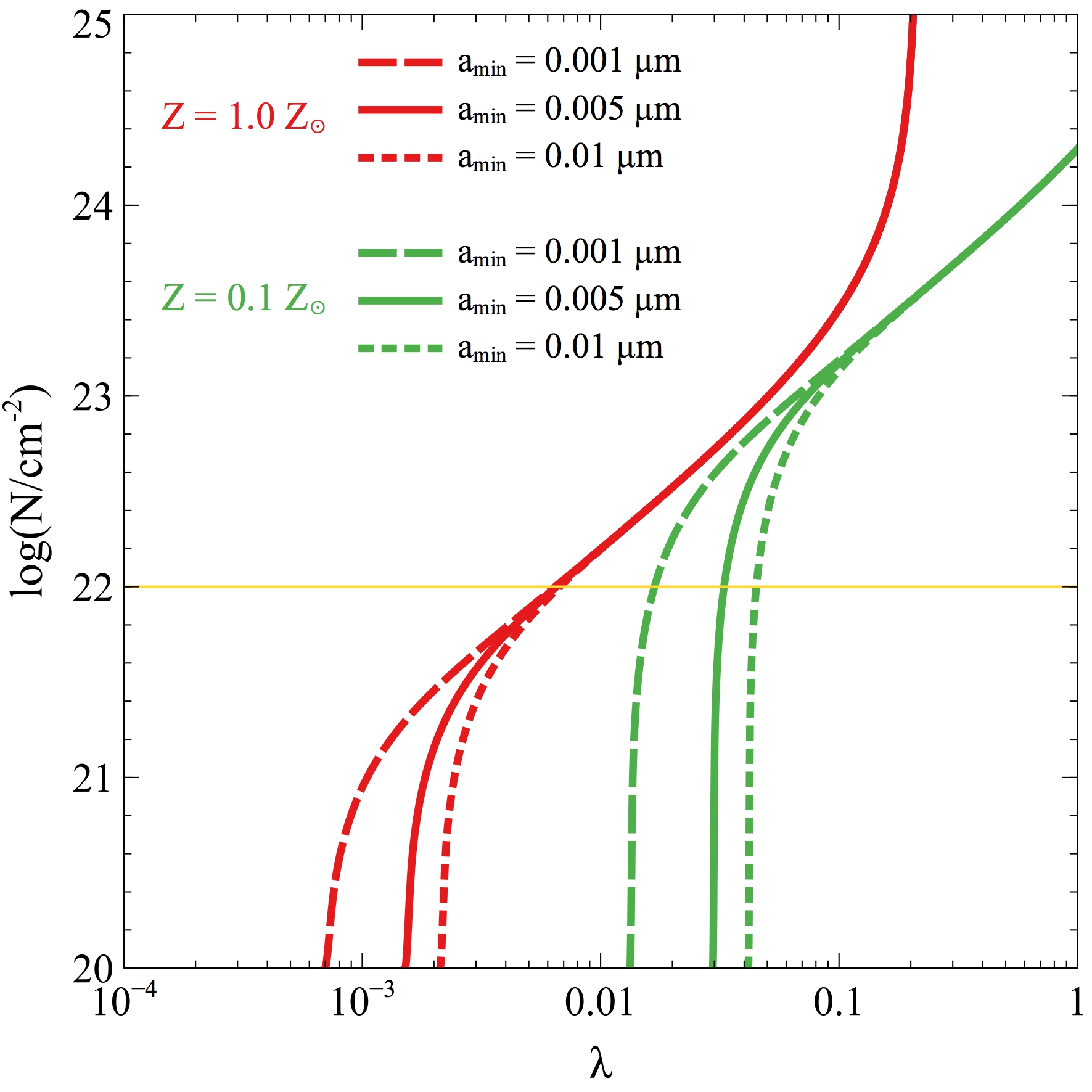}\par
\includegraphics[width=0.8\linewidth]{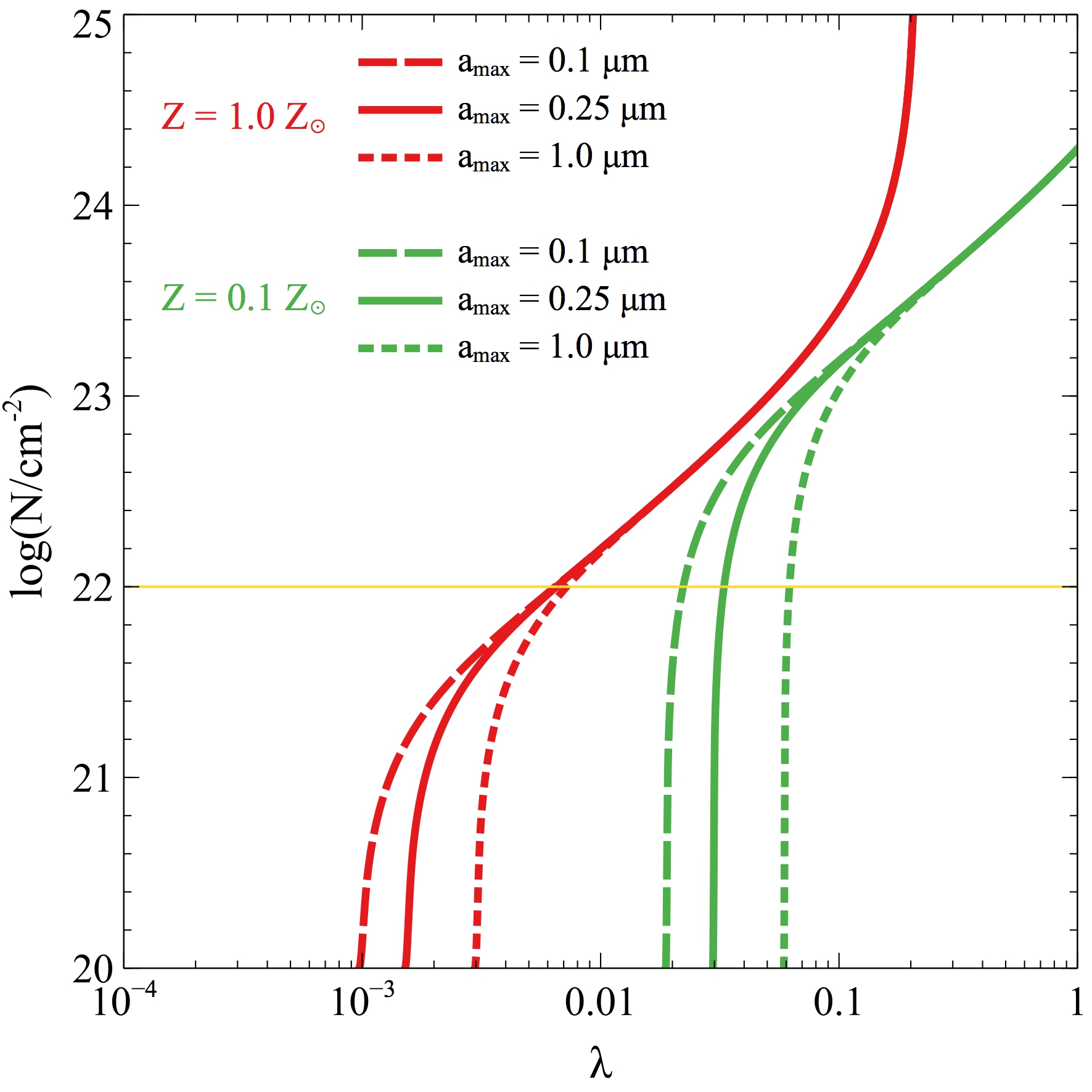}\par 
\end{multicols}
\caption{ 
$N_\mathrm{H} - \lambda$ plane with two metallicity cases: high-metallicity case ($Z = 1 Z_{\odot}$, red curves) and low-metallicity case ($Z = 0.1 Z_{\odot}$, green curves). 
Variations in minimum grain size (left-hand panel): $a_\mathrm{min} = 0.001 \mathrm{\mu m}$ (dashed), $a_\mathrm{min} = 0.005 \mathrm{\mu m}$ (solid), $a_\mathrm{min} = 0.01 \mathrm{\mu m}$ (dotted). 
Variations in maximum grain size (right-hand panel): $a_\mathrm{max} = 0.1 \mathrm{\mu m}$ (dashed), $a_\mathrm{max} = 0.25 \mathrm{\mu m}$ (solid), $a_\mathrm{max} = 1 \mathrm{\mu m}$ (dotted). 
}
\label{Fig_NH_lambda_var-amin-amax}
\end{figure*}

\begin{figure*}
\begin{multicols}{2} 
\includegraphics[width=0.8\linewidth]{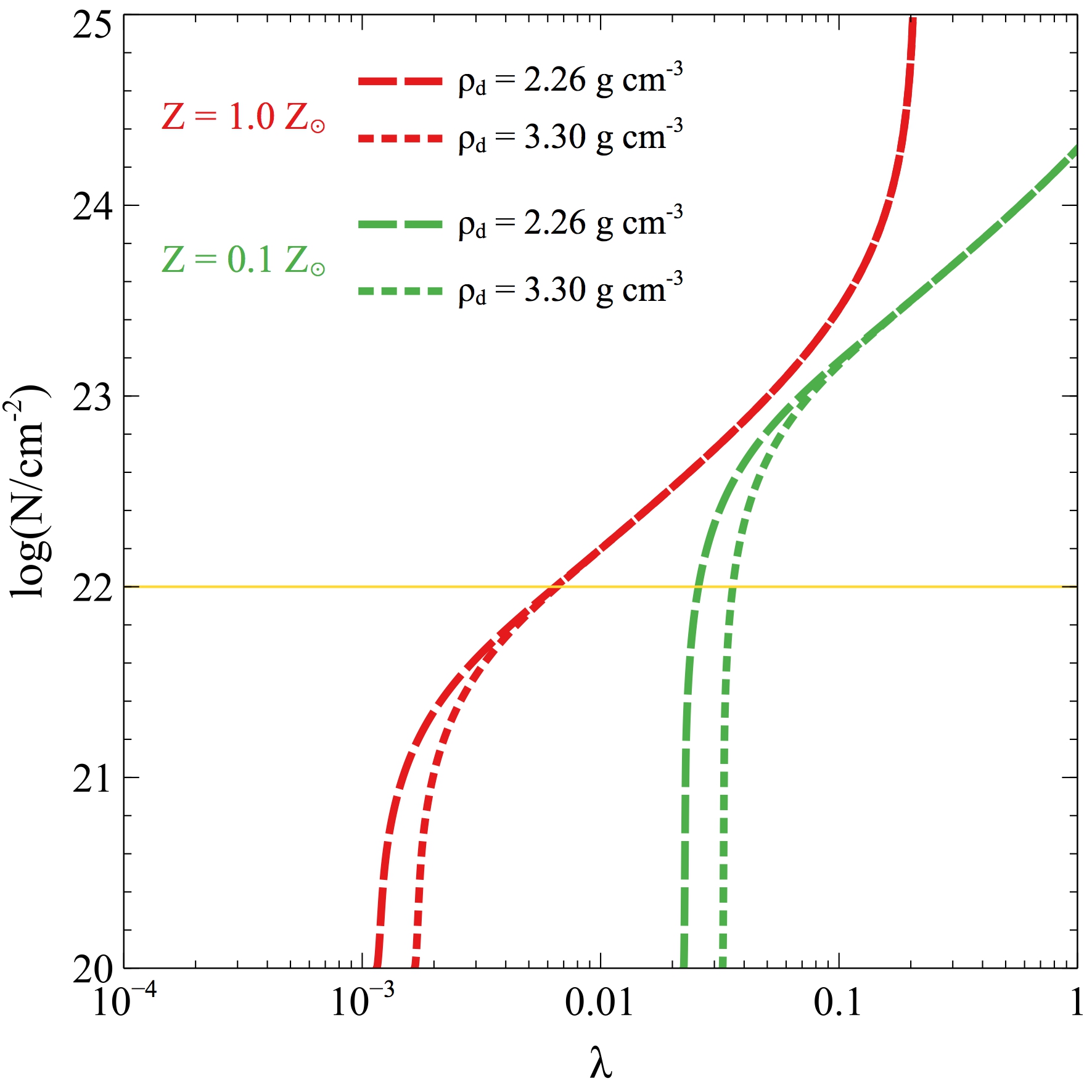}\par
\includegraphics[width=0.8\linewidth]{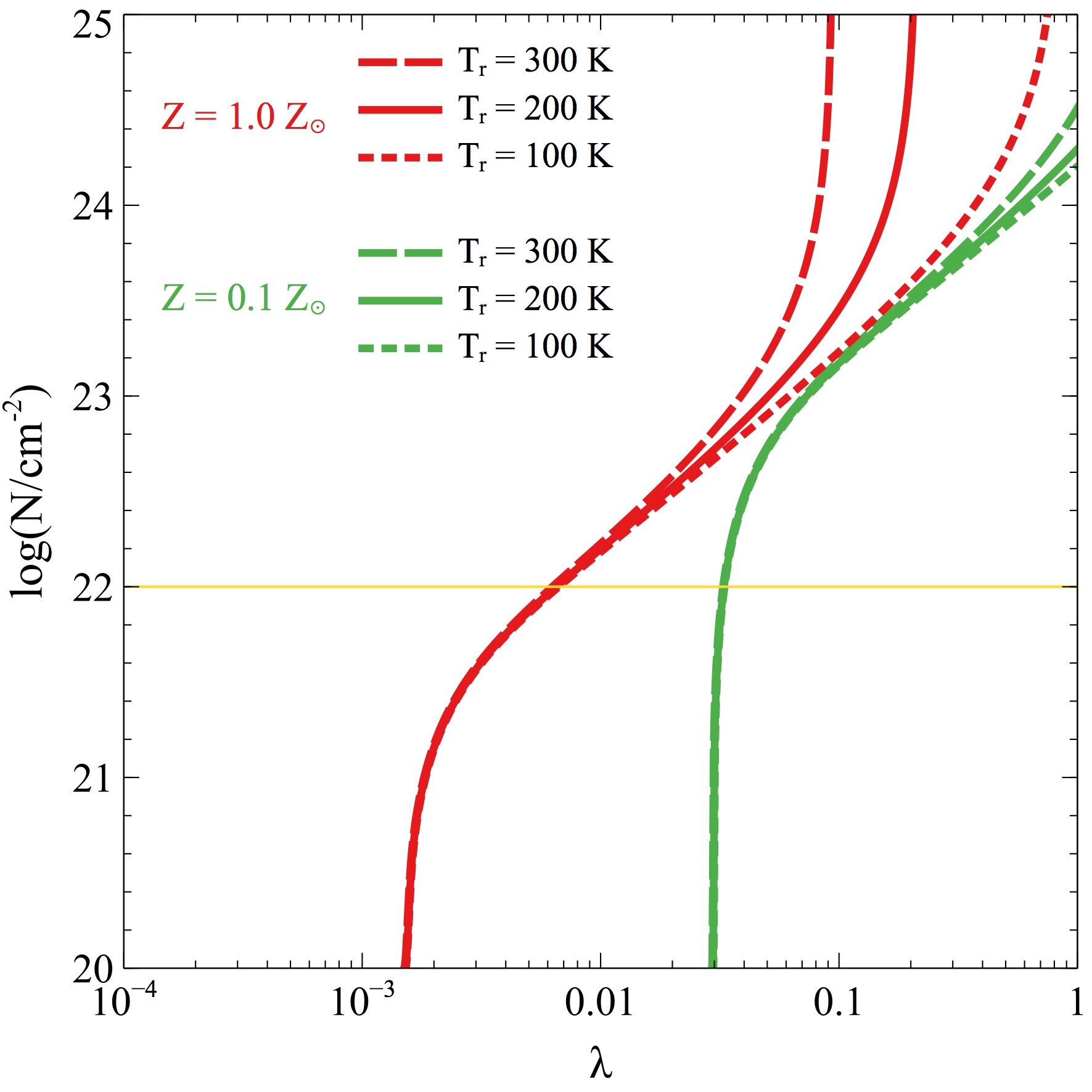}\par 
\end{multicols}
\caption{
$N_\mathrm{H} - \lambda$ plane with two metallicity cases: high-metallicity case ($Z = 1 Z_{\odot}$, red curves) and low-metallicity case ($Z = 0.1 Z_{\odot}$, green curves).
Variations in dust grain composition (left-hand panel): graphite grains $\rho_d = 2.26 \, \mathrm{g cm^{-3}}$ (dashed), and silicate grains $\rho_d = 3.3 \, \mathrm{g cm^{-3}}$ (dotted). 
Variations in radiation temperature (right-hand panel): $T = 300$ K (dashed), $T = 200 $ K (solid), $T = 100 $ K (dotted).
}
\label{Fig_NH_lambda_var-rho-T}
\end{figure*}


\subsubsection{Variations in dust grain size ($a_\mathrm{min}, a_\mathrm{max}$)}

We next analyse different grain size distributions by varying the minimum and maximum size of dust grains. Figure \ref{Fig_NH_lambda_var-amin-amax} shows the $N_\mathrm{H} - \lambda$ plane for two metallicities: a high-metallicity case (with solar value $Z = 1 Z_{\odot}$, red curves) and a low-metallicity case (with $Z = 0.1 Z_{\odot}$, green curves) for different values of $a_\mathrm{min}$ and $a_\mathrm{max}$. We see that variations in the grain size distribution can modify the boundary curves in the $N_\mathrm{H} - \lambda$ plane. We note that only the UV regime is affected: larger grains lead to lower UV opacities (due to the inverse square root dependence of $\kappa_\mathrm{UV} \propto a_\mathrm{min}^{-1/2} a_\mathrm{max}^{-1/2}$) and the boundary shifts to the right. As a result, the area of the forbidden region can be significantly reduced for larger grains, as evident in the low-metallicity case. 


\subsubsection{Variations in dust grain composition ($\rho_d$) }

Two different dust grain compositions ---graphite grains ($\rho_d = 2.26 \, \mathrm{g \, cm^{-3}}$) and silicate grains ($\rho_d = 3.3 \, \mathrm{g \, cm^{-3}}$)--- are considered in Fig. \ref{Fig_NH_lambda_var-rho-T} (left-hand panel). 
We observe that only the UV regime is affected, due to the inverse dependence of the UV opacity on grain density ($\kappa_\mathrm{UV} \propto 1/\rho_d$). The resulting difference is minor in the $N_\mathrm{H} - \lambda$ plane for the two types of grains considered here, with just a slight rightward shift of the boundary curve for denser silicate grains compared to lower density graphite grains. 


\subsubsection{Variations in radiation temperature ($T_r$)}

Finally, Figure \ref{Fig_NH_lambda_var-rho-T} (right-hand panel) shows the $N_\mathrm{H} - \lambda$ plane for different radiation temperatures. In this case, only the IR regime is affected, due to the quadratic dependence of the IR opacity on temperature ($\kappa_\mathrm{IR} \propto T_r^2$). Lower radiation temperatures shift the boundaries to the right in the $N_\mathrm{H} - \lambda$ plane, shrinking the area of the forbidden region at high column densities. As can be seen from the plot, this effect is prominent in the high-metallicity case, while it is much less noticeable in the low-metallicity case. 


\subsubsection{Other parameters}

In addition, the inclusion of the total enclosed mass (e.g. due to intervening stars or a nuclear star cluster) may further shift the boundary curve to the right \citep{Fabian_et_2009, Ishibashi_et_2018b}. 
Besides, the central black hole mass and spin determine the shape of the AGN spectral energy distribution, which in turn may affect the $N_\mathrm{H} - \lambda$ plane, although to a minor extent \citep{Arakawa_et_2022}. 


\section{X-ray weak JWST-AGN in the $N_\mathrm{H} - \lambda$ plane perspective}
\label{Sect_JWST_AGN}

We now consider JWST-AGN in the framework of the $N_\mathrm{H} - \lambda$ plane --by assuming a physically motivated choice for the parameters of the dusty gas. 


\subsection{The $N_\mathrm{H} - \lambda$ plane with low metallicity and little dust content}

Early JWST-AGN are hosted in metal-poor environments, with significantly sub-solar metallicities \citep{Harikane_et_2023, Kocevski_et_2023}. A gas phase metallicity of $Z \sim 0.2 Z_{\odot}$ is measured in a type 1.8 AGN at z = 5.55 \citep{Uebler_et_2023}. Similarly, a very low metallicity of $Z < 0.1 Z_{\odot}$ is derived in a spectroscopically confirmed LRD at z = 8.6 \citep{Tripodi_et_2024}. Low metallicities imply low dust-to-gas ratios, hence little dust content is expected in JWST-AGN at early times. The absorbing gas is indeed inferred to be extremely dust-poor in a high-redshift LRD at $z \sim 7$ \citep{DEugenio_et_2025}. 

The dust properties of early JWST-AGN could differ significantly from what observed in our Galaxy. Dust production in the first few hundred million years of the Universe is likely dominated by core-collapse supernovae, but the newly formed dust can be destroyed in reverse shocks, leading to significant uncertainties in the effective supernova dust yield \citep{Schneider_Maiolino_2024}. Small dust grains are preferentially destroyed by thermal or kinetic sputtering, while larger grains tend to better survive the supernova shocks \citep[][]{Tazaki_Ichikawa_2020, McKinney_et_2025}. On the other hand, grain growth can occur in the interstellar medium, and large dust grains may also form as a result of grain coagulation in the dense circumnuclear environment of AGNs \citep{Maiolino_et_2001b}. Despite the observational progress, there are several model assumptions and important uncertainties involved in the dust production/destruction mechanisms in the early Universe \citep[see the recent review by][and references therein]{Schneider_Maiolino_2024}.

\begin{figure}
\begin{center}
\includegraphics[angle=0,width=0.4\textwidth]{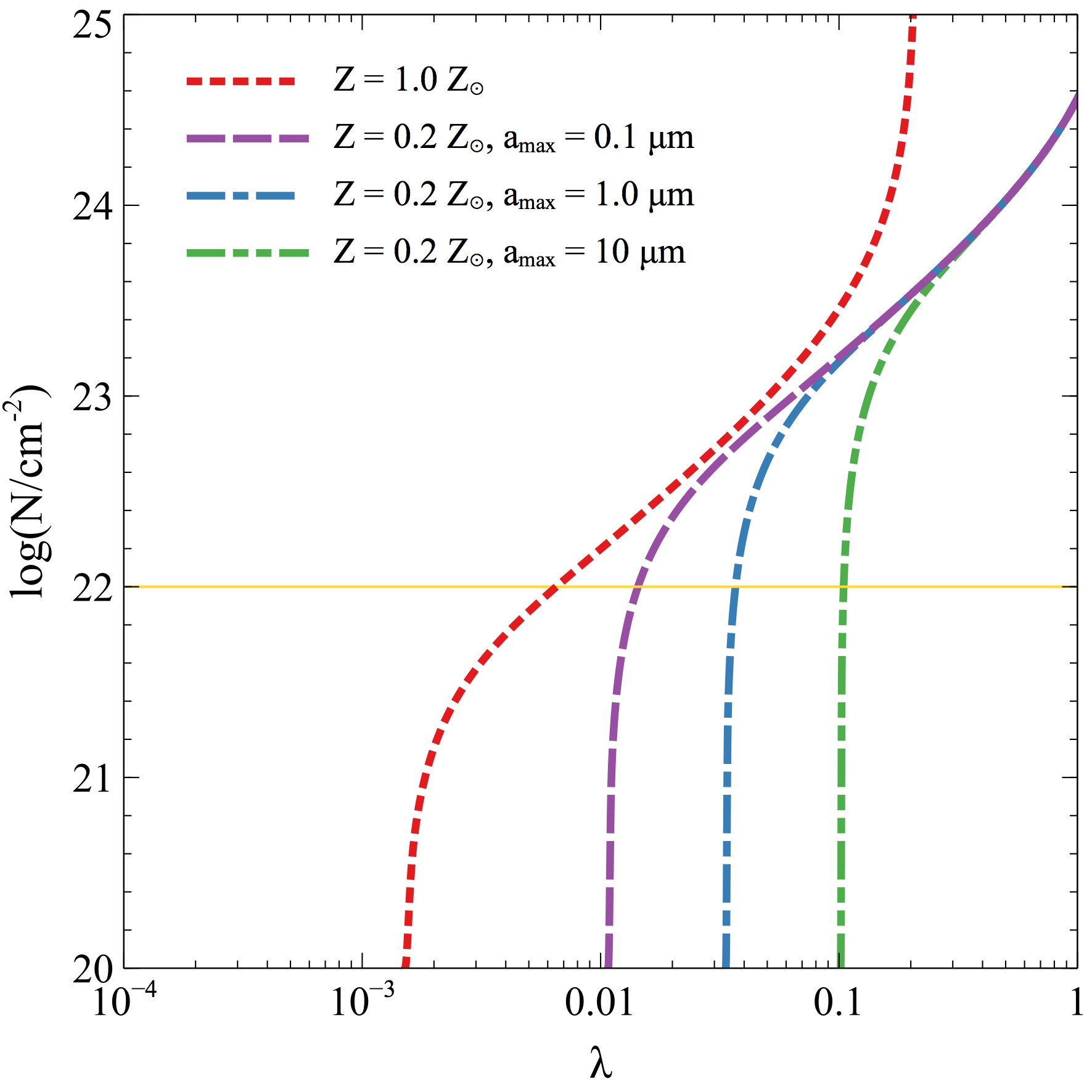} 
\caption{ 
$N_\mathrm{H} - \lambda$ plane for the solar metallicity case with fiducial parameters (red dotted) and JWST-AGN-like cases with lower metallicity ($Z = 0.2 Z_{\odot}$) and larger grains ($a_\mathrm{min} = 0.01 \mathrm{\mu m}$). Variations in maximum grain size: $a_\mathrm{max} = 0.1 \mathrm{\mu m}$ (violet dashed), $a_\mathrm{max} = 1.0 \mathrm{\mu m}$ (blue dash-dot), $a_\mathrm{max} = 10 \mathrm{\mu m}$ (green dash-dot-dot).   
}
\label{Fig_NH_lambda_JWST}
\end{center}
\end{figure}

We assume a gaseous medium with low metallicity and large dust grains as a representative case of JWST-AGN in the early cosmic epoch. Figure \ref{Fig_NH_lambda_JWST} shows the $N_\mathrm{H} - \lambda$ plane, comparing the solar metallicity case ($Z = 1 Z_{\odot}$, with standard dust parameters) to the low metallicity cases ($Z = 0.2 Z_{\odot}$, with larger dust grains). We observe a significant difference between the two metallicity cases; for a given (low) metallicity, larger grains further reduce the area of the forbidden region --- or equivalently enhance the region of long-lived obscuration. 

Physically, a combination of low metallicity (i.e. low dust-to-gas ratio) and large grains decrease the UV dust opacity, which shifts the boundary curve to the right in the $N_\mathrm{H} - \lambda$ plane. As a consequence, higher Eddington ratios ($\lambda$) will be required to blow out a given column density ($N$) of dusty gas. Conversely, dust-poor gas can survive against radiation pressure (even at moderately high Eddington ratios) and pile up in the nuclear region. Such long-lived absorbing gas may be at the origin of heavy X-ray obscuration.     

\subsection{Dust extinction and the $A_\mathrm{V} - \lambda$ plane}

The $N_\mathrm{H} - \lambda$ plane may be converted into a corresponding $A_\mathrm{V} - \lambda$ plane, by assuming a certain $A_\mathrm{V} - N_\mathrm{H}$ relation between the two quantities \citep[e.g.][]{Mizukoshi_et_2024}. From the definition of the total-to-selective extinction ratio $R_\mathrm{V} = A_\mathrm{V}/E_\mathrm{B-V}$ (where $A_\mathrm{V}$ is the V-band dust extinction and $E_\mathrm{B-V}$ is the dust reddening), one derives $A_\mathrm{V} = R_\mathrm{V} E_\mathrm{B-V} = R_\mathrm{V} \frac{E_\mathrm{B-V}}{N_\mathrm{H}} N_\mathrm{H}$. 

The $(E_\mathrm{B-V}/N_\mathrm{H})$ ratio in local AGNs is found to be lower by a factor of $\sim 3 - 100$ \citep{Maiolino_et_2001b}, compared to the standard Galactic value of $(E_\mathrm{B-V}/N_\mathrm{H})_\mathrm{Gal} = 1.7 \times 10^{-22} \, \mathrm{mag \, cm^2}$ \citep{Bohlin_et_1978}. Similarly reduced $(E_\mathrm{B-V}/N_\mathrm{H})$ ratios have also been observed in other obscured AGNs and dust-reddened quasars at high redshifts \citep{Jun_et_2021, Glikman_et_2024}.  
The low values of the $(E_\mathrm{B-V}/N_\mathrm{H})$ ratio could be in part attributed to the presence of large dust grains in the AGN nuclear region \citep{Maiolino_et_2001b}. 

A prevalence of larger grains, coupled with a lack of smaller grains, lead to flatter extinction curves. Flat extinction curves have been reported for AGNs \citep{Gaskell_et_2004}, with $R_\mathrm{V}$ ratios larger than the canonical value of $R_\mathrm{V} = 3.1$. 
Supernova dust --characterised by a dominance of large grains ($\gtrsim 0.1 \mathrm{\mu m}$) and a deficit of small grains ($\lesssim 0.01 \mathrm{\mu m}$)-- may also yield a flat dust attenuation curve at early cosmic epochs \citep[][]{Markov_et_2025}. It has even been argued that grey extinction curves may explain the V-shaped spectral energy distribution of LRDs \citep{Killi_et_2024, Li_et_2025}. 

\begin{figure}
\begin{center}
\includegraphics[angle=0,width=0.4\textwidth]{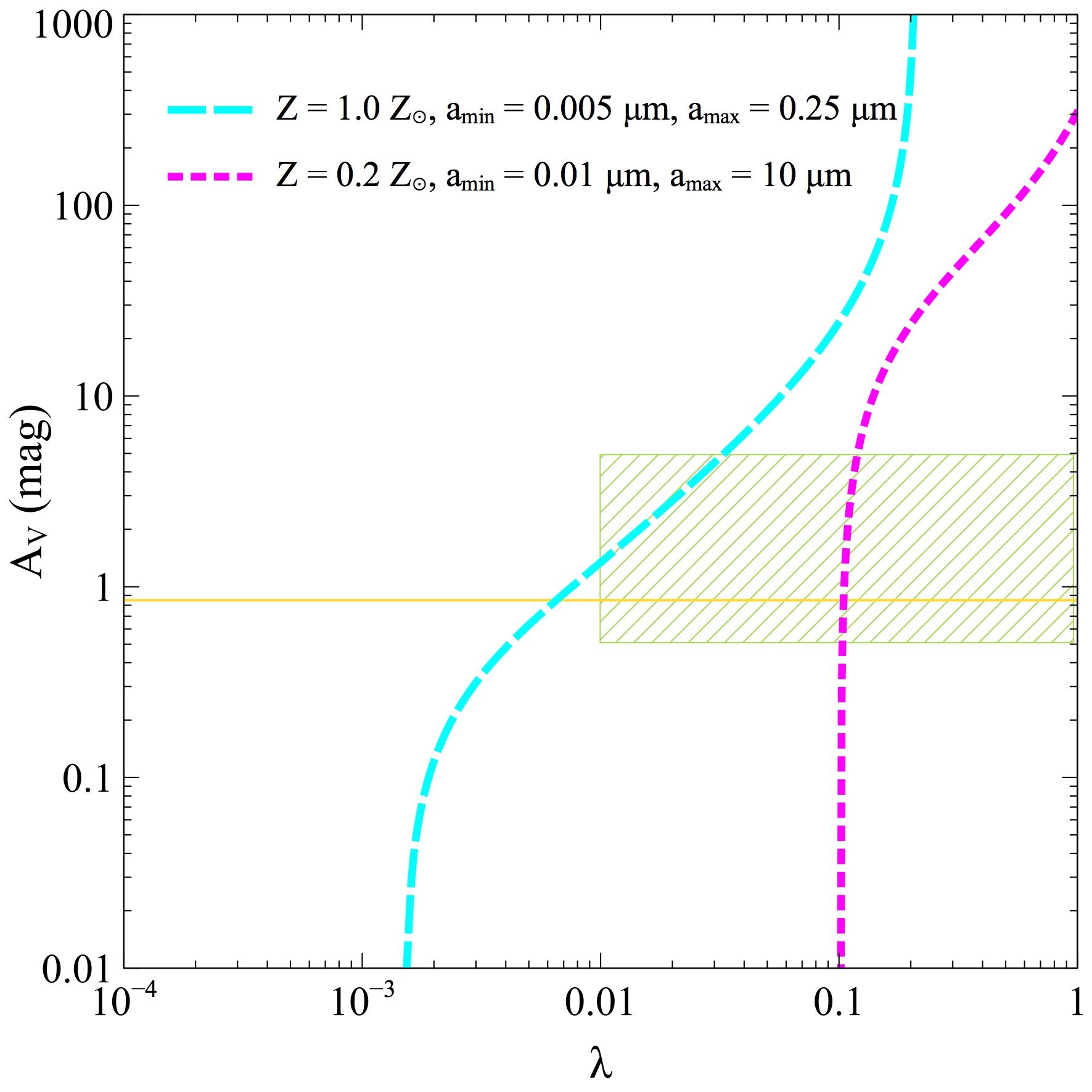}
\caption{ 
$A_\mathrm{V} - \lambda$ plane with $R_\mathrm{V} = 5.0$ and $E_\mathrm{B-V}/N_\mathrm{H} = 0.1 \left(E_\mathrm{B-V}/N_\mathrm{H} \right)_\mathrm{Gal}$. 
Solar metallicity case with fiducial parameters (cyan dashed) and JWST-AGN-like case with lower metallicity ($Z = 0.2 Z_{\odot}$) and larger dust grains ($a_\mathrm{min} = 0.01 \mathrm{\mu m}$, $a_\mathrm{max} = 10 \mathrm{\mu m}$) (magenta dotted). The horizontal line (yellow fine line) corresponds to the $N = 10^{22} \mathrm{cm^{-2}}$ limit. Rectangular shade (green hatched): $A_\mathrm{V} \sim (0.5 - 5)$ mag and $\lambda \sim (0.01 - 1)$.  
}
\label{Fig_AV_lambda}
\end{center}
\end{figure}

Assuming a flat extinction curve ($R_\mathrm{V} = 5.0$) and a reduced ratio of $(E_\mathrm{B-V}/N_\mathrm{H}) \sim 0.1 (E_\mathrm{B-V}/N_\mathrm{H})_\mathrm{Gal}$, we obtain $A_\mathrm{V} = 8.5 \times 10^{-23} N_\mathrm{H}$. 
Figure \ref{Fig_AV_lambda} shows the resulting $A_\mathrm{V} - \lambda$ plane for two cases: a solar metallicity case with standard dust properties (cyan curve) and a low-metallicity case with large dust grains (magenta curve) typical of JWST-AGN.  
A certain amount of dust extinction and reddening is observed in most type-1 JWST-AGN \citep{Maiolino_et_2025}.
In fact, JWST observations of LRDs indicate low to moderate dust extinction, typically in the range $A_\mathrm{V} \sim 0.5 - 5$ mag \citep[][and references therein]{Casey_et_2024}, represented by a rectangular shade in Fig. \ref{Fig_AV_lambda} (for illustrative purposes). While most of the sources in the shaded area should be outflowing at solar metallicity, in the low-metallicity case, a significant fraction would be located in the long-lived obscuration region. 

A mismatch between optical obscuration and X-ray absorption has been previously observed in AGNs \citep{Burtscher_et_2016, Liu_et_2018, Shimizu_et_2018}. One possible explanation is heavy X-ray absorption taking place in high-density dust-free gas, e.g. within the broad line region. In fact, BLR clouds usually have high column densities ($\gtrsim 10^{24} \mathrm{cm^{-2}}$) and are naturally devoid of dust, being located inside the dust sublimation radius. These Compton-thick dust-free clouds along the line of sight may account for the high values of the $N_\mathrm{H}/A_\mathrm{V}$ ratio. 

\section{AGN obscuring clouds}
\label{Sect_Clouds}

\subsection{Blowout vs. stalling conditions}

The obscuring material may be in the form of clouds or clumps distributed in a quasi-spherical configuration around the central AGN. 
The equation of motion of a cloud of mass $M_c$ and radius $R_c$ is given by \citep[cf.][]{Ishibashi_et_2024}
\begin{equation}
\frac{d}{dt} \left[ M_c v_c \right] = \frac{L}{c} \left( \frac{ \pi R_c^2}{4 \pi r^2} \right) \left[  \tau_\mathrm{IR} + 1 - e^{-\tau_\mathrm{UV}} \right] - \frac{G M_\mathrm{BH} M_c}{r^2} , 
\label{Eq_motion_clump}
\end{equation} 
where $\tau_\mathrm{IR,UV} = \frac{\kappa_\mathrm{IR, UV} M_c}{\pi R_c^2} = \kappa_\mathrm{IR,UV} m_p N_c$, and $N_c$ is the clump column density. 
As in section \ref{Sect_NH_lambda}, we assume a predominantly neutral gas.  
Equating the radiative and gravitational forces, we obtain the effective Eddington luminosity for the cloud, $L_\mathrm{E,c}^{'}$. 
The corresponding effective Eddington ratio of the clump can be written as 
\begin{equation}
\Lambda_c = \frac{L}{L_\mathrm{E,c}^{'}} = \frac{L R_c^2( \tau_\mathrm{IR} + 1 - e^{-\tau_\mathrm{UV}})}{4 c G M_\mathrm{BH} M_c} 
=  \frac{\lambda( \tau_\mathrm{IR} + 1 - e^{-\tau_\mathrm{UV}})}{\sigma_T N_c} , 
\end{equation}
where we have introduced the classical Eddington ratio $\lambda = L/L_\mathrm{E}$ in the last equality. 
We recall that dusty gas is ejected when the effective Eddington limit is exceeded, i.e. when the effective Eddington ratio exceeds unity. The blow-out condition for the obscuring cloud is thus given by $\Lambda_c > 1$. 

\begin{figure*}
\begin{multicols}{2}
\includegraphics[width=0.8\linewidth]{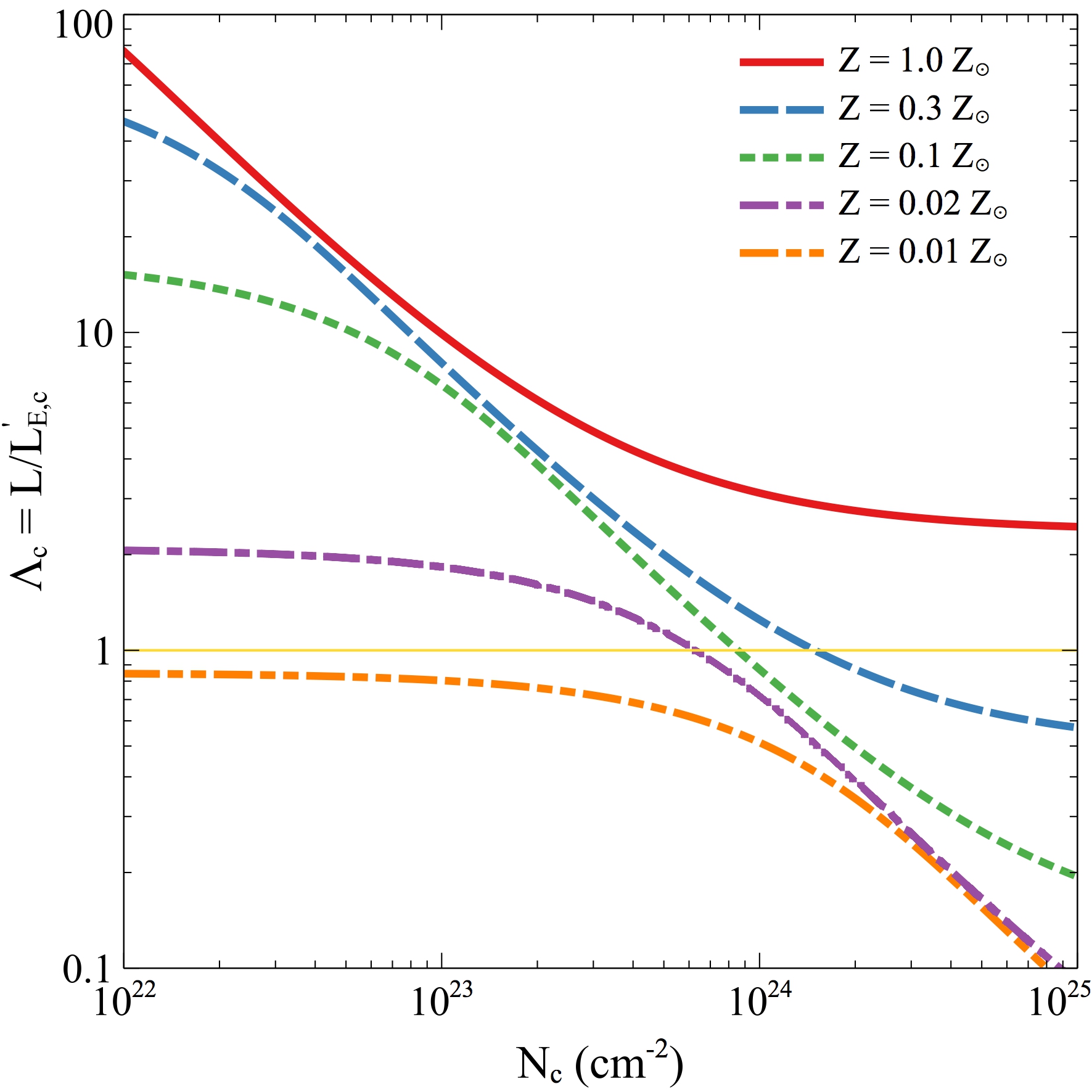}\par
\includegraphics[width=0.8\linewidth]{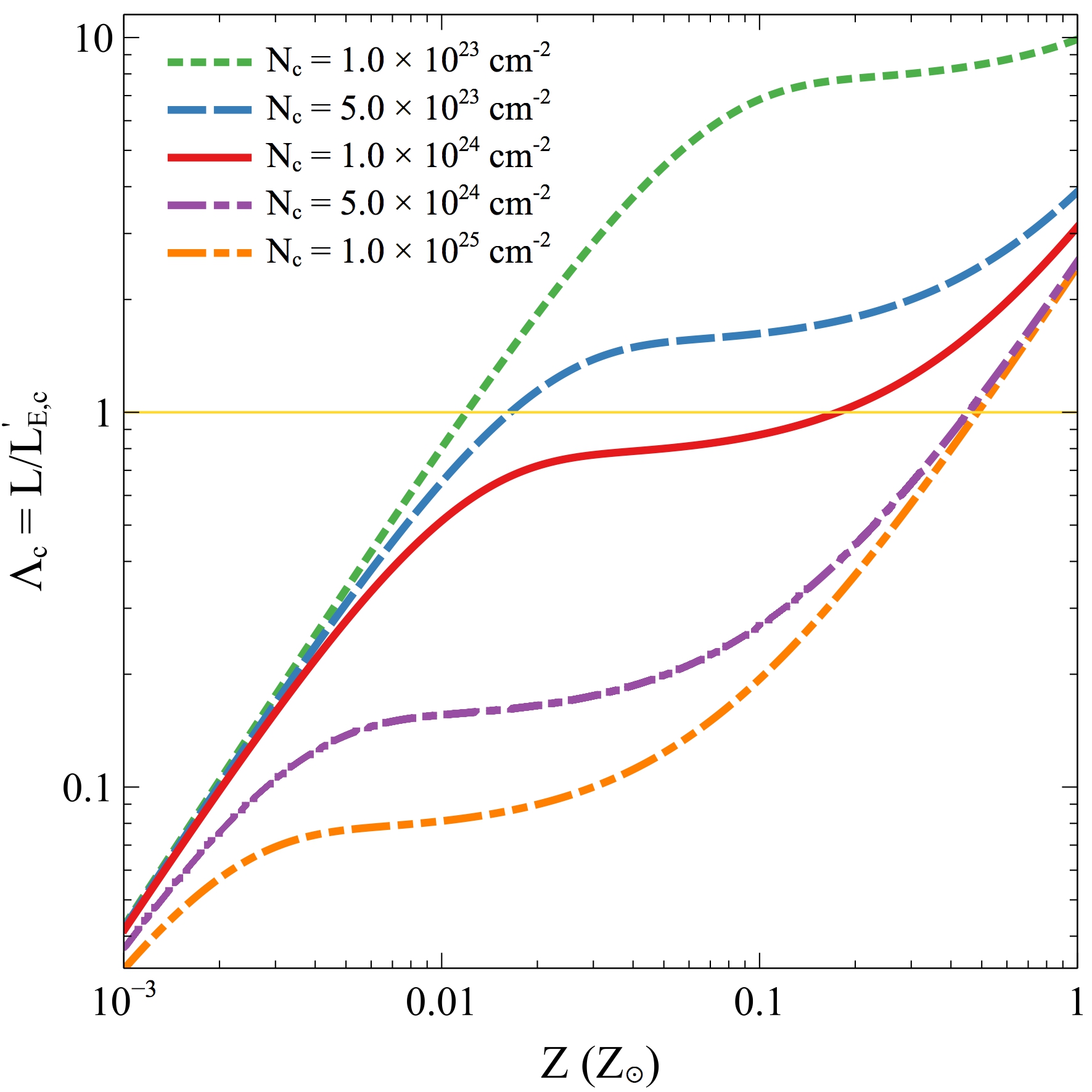}\par  
\end{multicols}
\caption{
Left-hand panel: effective Eddington ratio ($\Lambda_c$) as a function of clump column density ($N_c$) for $\lambda = 0.5$ and different metallicities: $Z = 1 Z_{\odot}$ (red solid), $Z = 0.3 Z_{\odot}$ (blue dashed), $Z = 0.1 Z_{\odot}$ (green dotted), $Z = 0.02 Z_{\odot}$ (violet dash-dot), $Z = 0.01 Z_{\odot}$ (orange dash-dot-dot). 
Right-hand panel: effective Eddington ratio ($\Lambda_c$) as a function of metallicity ($Z$) for $\lambda = 0.5$ and different clump column densities: $N_c = 10^{23} \mathrm{cm^{-2}}$ (green dotted), $N_c = 5 \times 10^{23} \mathrm{cm^{-2}}$ (blue dashed), $N_c = 10^{24} \mathrm{cm^{-2}}$ (red solid), $N_c = 5 \times 10^{24} \mathrm{cm^{-2}}$ (violet dash-dot), $N_c = 10^{25} \mathrm{cm^{-2}}$ (orange dash-dot-dot). 
The horizontal line (yellow fine line) marks the blowout condition $\Lambda_c = 1$ in both panels. 
}
\label{Fig_Lambda_Nc_Z}
\end{figure*}

In Fig. \ref{Fig_Lambda_Nc_Z} (left-hand panel), we plot the effective Eddington ratio ($\Lambda_c$) as a function of the clump column density ($N_c$) for different values of the metallicity ($Z$), at a fixed $\lambda = 0.5$.
 In general, the effective Eddington ratio decreases with increasing clump column density, while higher metallicities lead to higher effective Eddington ratios. In the case of solar metallicity ($Z = 1 Z_{\odot}$, red curve), the effective Eddington ratio is always greater than unity ($\Lambda_c > 1$); whereas for $Z = 0.01 Z_{\odot}$ (orange curve), the effective Eddington ratio stays below unity ($\Lambda_c < 1$) at all $N_c$. This implies that obscuring clouds cannot be blown out at very low metallicities because the blow-out condition is not satisfied. At intermediate metallicities ($Z/Z_{\odot} \sim 0.02 - 0.3$), the effective Eddington ratio drops below unity above a critical clump column density ($N_\mathrm{c,max}$), which increases with increasing metallicity. In fact, higher metallicities are required to blow out heavier columns (with $N_\mathrm{c,max}$ corresponding to the maximal column density that can be ejected at a given metallicity).  

A complementary view is provided in Fig. \ref{Fig_Lambda_Nc_Z} (right-hand panel), which shows the effective Eddington ratio as a function of metallicity, for different values of the clump column density, again at $\lambda = 0.5$. We see that the effective Eddington ratio generally increases with increasing metallicity, and higher $\Lambda_c$ are obtained for lower clump column densities. The effective Eddington ratio exceeds unity above a critical metallicity threshold ($Z_\mathrm{c,min}$), implying that a minimal metallicity value is required to clear out a given column density. The critical metallicities are higher for heavier column densities, i.e. higher metallicities are required to eject heavier columns. In other words, low metallicities allow large column densities of gas to accumulate in the nuclear region, leading to long-lived obscuration. 

\subsection{X-ray spectral simulations}

The escaping X-ray flux can be severely reduced by absorption and scattering processes at high column densities. 
At parsec scales, the most important of these are Compton scattering, photoelectric absorption from the K shell of metal species, and subsequent fluorescent line emission. We carry out Monte-Carlo calculations for a cold, neutral, uniform spherical cloud surrounding an emitting source with power law index of $\Gamma$ = 2 (Gursahani $\&$ Reynolds, in preparation). The radius of the cloud is 0.75 pc, but this does not affect the results. Since we have a uniform medium, the column density uniquely determines the optical depth. Compton scattering is treated fully, with electron differential cross-sections determined as a function of photon energy. We include K-shell absorption and K$\alpha$ line emission from 11 elements: C, O, Ne, Mg, Si, S, Ar, Ca, Cr, Fe, and Ni. Additionally, we split the Fe K$\alpha$ line into K$\alpha_1$ and K$\alpha_2$ and add the K$\beta$ line for iron as well. 

\begin{figure*}
\begin{multicols}{2}
\includegraphics[width=1.0\linewidth]{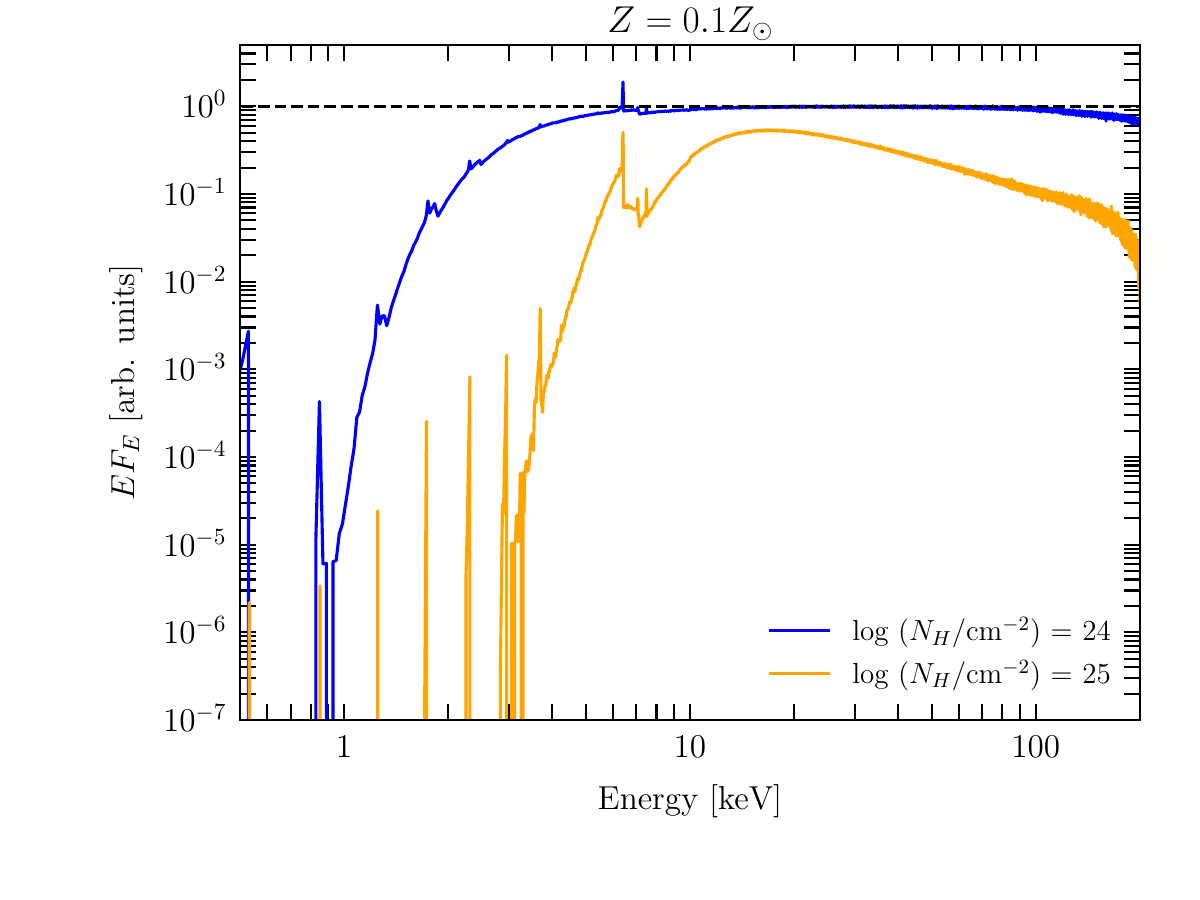}\par
\includegraphics[width=1.0\linewidth]{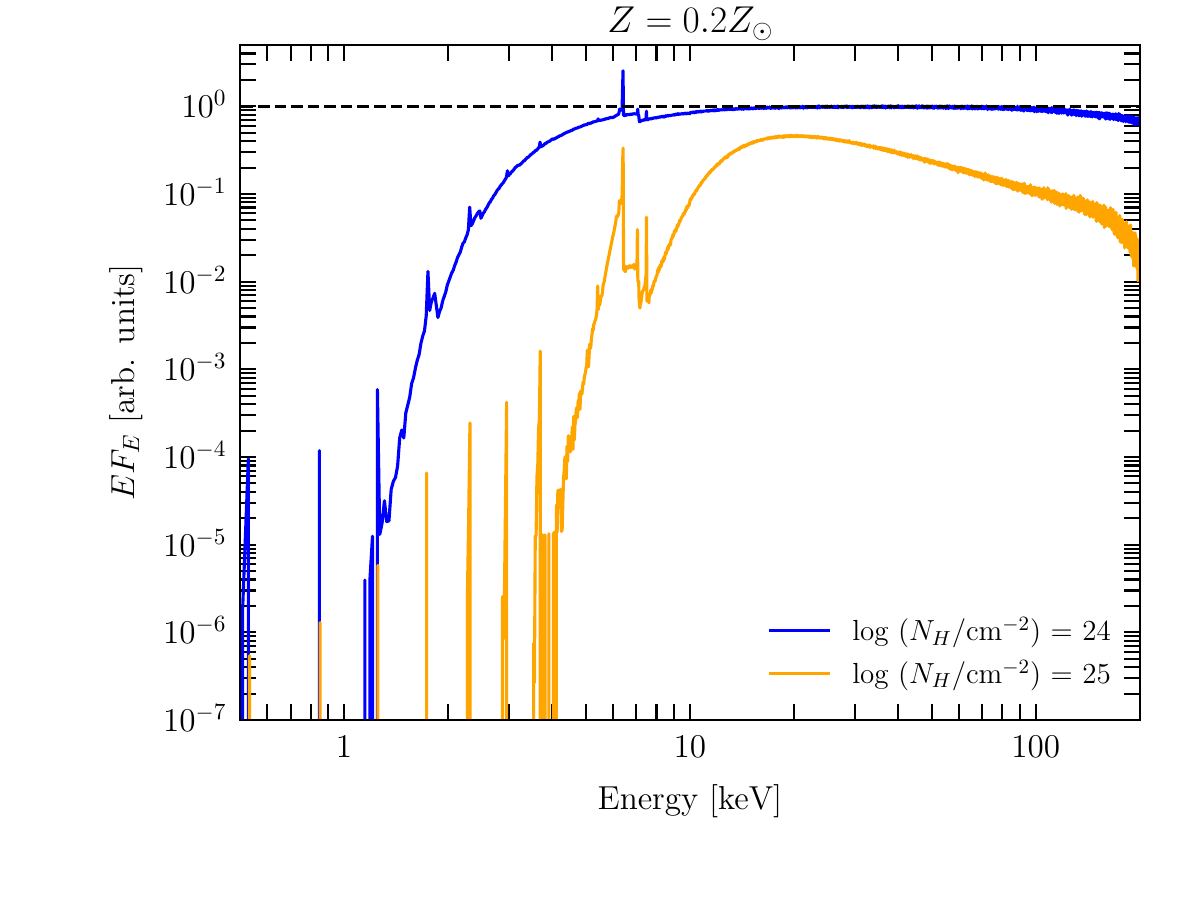}\par  
\end{multicols}
\caption{
Simulated X-ray spectra for a spherical cloud with $Z = 0.1 Z_{\odot}$ (left-hand panel) and $Z = 0.2 Z_{\odot}$ (right-hand panel), with $N_\mathrm{H} = 10^{24} \mathrm{cm^{-2}}$ (blue curve) and $N_\mathrm{H} = 10^{25} \mathrm{cm^{-2}}$ (orange curve). The black dashed line is the input power-law. 
}
\label{Fig_X_spectra}
\end{figure*}

Figure \ref{Fig_X_spectra} shows the resulting simulated X-ray spectra for $Z = 0.1 Z_{\odot}$ (left-hand panel) and $Z = 0.2 Z_{\odot}$ (right-hand panel). The corresponding values of the absorbed flux fractions and photon fractions are tabulated in Table \ref{Table_X_spectra}. As expected, the X-ray flux can be strongly absorbed, with enhanced effects at higher column densities and higher metallicities. 

\begin{table}
\centering
\renewcommand{\arraystretch}{1.4}
\begin{tabular}{|c|c|c|}
\hline
\textbf{Z} & \textbf{log (N$_\text{H}$/cm$^{-2}$) = 24} & \textbf{log (N$_\text{H}$/cm$^{-2}$) = 25} \\
\hline
\textbf{0.1 Z$_\odot$} & 
\begin{tabular}[c]{@{}l@{}}
2-20 keV: 0.137 (0.439)\\
10-70 keV: 0.021 (0.022)
\end{tabular} & 
\begin{tabular}[c]{@{}l@{}}
2-20 keV: 0.706 (0.906)\\
10-70 keV: 0.680 (0.458)
\end{tabular} \\
\hline
\textbf{0.2 Z$_\odot$} & 
\begin{tabular}[c]{@{}l@{}}
2-20 keV: 0.215 (0.591)\\
10-70 keV: 0.028 (0.043)
\end{tabular} & 
\begin{tabular}[c]{@{}l@{}}
2-20 keV: 0.804 (0.951)\\
10-70 keV: 0.709 (0.591)
\end{tabular} \\
\hline
\end{tabular}
\caption{Absorbed flux fractions ($EF_E$) with corresponding absorbed photon fractions ($N_E$) in parentheses for the 2-20 keV and 10-70 keV in the rest frame of the source. These fractions are calculated with the Monte Carlo code described in the text. We use a power-law index $\Gamma = 2$ for all input spectra.}
\label{Table_X_spectra}
\end{table}


\section{ Discussion }
\label{Sect_Discussion}


\subsection{ The X-ray weakness problem in JWST-AGN }

One of the unexpected observational results is the apparent lack of X-ray emission from early JWST-AGN. X-ray observations of individual LRDs result in non-detection, and stacking analysis only yield a tentative detection, suggesting that LRDs are X-ray weak \citep{Ananna_et_2024, Yue_et_2024}. A more general investigation of the bulk JWST-AGN population (not restricted to LRDs) indicates that the great majority is X-ray undetected, with a non-detection even in the case of stacked X-ray data in the deepest Chandra fields \citep{Maiolino_et_2025}. 

The X-ray upper limits are $\sim 1-2$ orders of magnitude below what would be expected from a typical AGN spectral energy distribution. Compared to the local relation between X-ray and bolometric luminosities (the $L_\mathrm{X} - L_\mathrm{bol}$ relation), very high values of the X-ray bolometric correction ($K_\mathrm{X} = L_\mathrm{bol}/L_\mathrm{X}$) are derived for JWST-AGN. 
For instance, GN-28074 which is a broad-line AGN at $z = 2.26$, has the highest ratio of $L_\mathrm{bol}/L_\mathrm{X} > 10^4$, and is one of the most X-ray weak AGN known \citep{Juodzbalis_et_2024}. 

Two physical scenarios have been debated in the literature to explain the X-ray weakness of JWST-AGN. One possible solution is super-Eddington accretion giving rise to AGN spectral energy distribution with intrinsically weak X-ray emission. The super-Eddington models are extensively discussed in other works \citep{Pacucci_Narayan_2024, Madau_Haardt_2024, King_2025}.

An alternative possibility is heavy X-ray absorption by high-column density, likely Compton-thick gas \citep{Maiolino_et_2025, Juodzbalis_et_2024}. Since most type-1 JWST-AGN are X-ray obscured, the absorbing medium should be dust-poor. It has been suggested that dense BLR clouds, which are intrinsically dust-free, may be responsible for X-ray absorption.\footnote{Moderate dust reddening is actually observed in JWST-AGN, suggesting that some absorption may also occur outside the BLR.} The covering factor of BLR clouds must be very high, close to unity, in order to account for the very high fraction of X-ray undetected JWST-AGN. Such large covering factors may be expected when dense gas lingers in the nuclear region, due to weak AGN radiative feedback. 
A common picture emerging from the latest JWST observations is that of a central black hole enshrouded in a dense gaseous cocoon \citep{Rusakov_et_2025, Naidu_et_2025, deGraaff_et_2025, Ji_et_2025, DEugenio_et_2025}. 

We have seen that low metallicities imply low dust content, hence reduced radiation pressure feedback (Sections \ref{Sect_NH_lambda}-\ref{Sect_JWST_AGN}). AGN radiative feedback is unable to clear out dust-poor gas because, in the near-absence of dust, the radiation-matter coupling is considerably reduced. As a consequence, the dust-poor gas cannot be easily removed and accumulates in the nuclear region. The absorbing gas is often observed to be at rest close to the BLR, as indicated by JWST spectroscopy \citep{DEugenio_et_2025}. Such rest-frame absorbers with little dust content likely form stationary structures that can be associated with stalling gas clouds. This may naturally account for the high covering fraction of X-ray absorbing clouds. From the observational viewpoint, the high covering factor of BLR clouds is also supported by the large equivalent widths of the broad $\mathrm{H \alpha}$ component observed in JWST-AGN \citep{Maiolino_et_2025}. 


\subsection{ Additional peculiar features of JWST-AGN }

In addition to X-ray weakness, a number of other peculiar properties are observed in JWST-AGN at early cosmic times. 

\subsubsection{Balmer absorption}

Balmer absorption features in the $\mathrm{H\alpha}$ and/or $\mathrm{H \beta}$ line profiles are observed in a significant fraction of JWST-AGN, around $\sim (10-20) \%$ \citep{Matthee_et_2024, Kocevski_et_2024, Lin_et_2024}. 
Balmer absorption lines also imply a Balmer break of non-stellar origin \citep{Inayoshi_Maiolino_2025, Ji_et_2025}. The presence of Balmer absorption requires very high gas densities ($> 10^9 \mathrm{cm^{-3}}$), whereby the $n = 2$ state is populated by collisionally excited hydrogen atoms. The high-density gas can be identified with the dense gas clouds in the BLR and its vicinity. These same BLR clouds may form the Compton-thick medium responsible for heavy X-ray absorption.  Interestingly, the few X-ray detected JWST-AGN have X-ray spectra consistent with being Compton-thick sources \citep[][and references therein]{Maiolino_et_2025}.

\subsubsection{Radio weakness }

Radio emission should provide a complementary view of JWST-AGN, as it is little affected by obscuration. Yet, radio observations of photometrically selected LRDs and spectroscopically confirmed broad-line AGNs yield non-detection, even in radio stacking analysis \citep{Mazzolari_et_2024b, Perger_et_2025, Gloudemans_et_2025}. This suggests that JWST-AGN are likely radio-quiet sources, for instance GN-28074 is extremely radio-weak \citep{Juodzbalis_et_2024}. A possible origin for the radio weakness is free-free absorption in a dense absorbing medium \citep{Mazzolari_et_2024b}. The dense BLR clouds that obscure the X-rays may also absorb radio emission --provided that the radio emission is confined within a region comparable to the BLR. In this picture, both X-ray weakness and radio weakness are physically connected and share a common origin in JWST-AGN. 

\subsubsection{Lack of ionized outflows}

Early JWST-AGN lack strong ionised outflows on galactic scales, as indicated by their narrow and symmetric [OIII] line profiles \citep{Maiolino_et_2025}. This contrasts with the broad [OIII] profiles (with prominent blueshifted wings tracing high-velocity winds) observed in lower redshift AGNs, as well as luminous quasars at high redshifts. The absence of strong outflows in JWST-AGN may be attributed to weak radiative dusty feedback, possibly due to low metallicity and low dust content. AGN radiation pressure is indeed unable to drive large-scale outflows in low-metallicity environments. As a consequence, dust-poor clouds cannot be efficiently evacuated from the circumnuclear region and further contribute to X-ray obscuration.  

\subsubsection{Faint FeII emission}

The FeII bump in the optical spectrum of JWST-AGN is found to be significantly weaker than in local AGN counterparts \citep{Trefoloni_et_2024}. The low values of the $R_\mathrm{FeII}$ ratio (defined by the ratio of the equivalent widths of FeII and $\mathrm{H\beta}$ lines) could be due to the low metallicity of the BLR in JWST-AGN. In contrast, high-luminosity sources at high redshifts have FeII emission of comparable strength to lower-redshift counterparts. It is interesting to note that the only two X-ray detected AGNs in the sample of \citet{Trefoloni_et_2024} tend to have the highest $R_\mathrm{FeII}$ values. Thus a connection between X-ray weakness and $R_\mathrm{FeII}$ weakness could also be envisaged, with both being driven by the low metal content. Therefore low metallicity may be a key parameter governing the peculiar properties of JWST-AGN. 


\subsection{Comparison to other AGN populations }

As we have seen, JWST-AGN present a number of peculiar features that set them apart from both low redshift AGNs and luminous quasars at high redshifts. For instance, early quasars possess large-scale ionised winds and display $R_\mathrm{FeII}$ ratios comparable to lower redshift counterparts. This is consistent with the fact that luminous quasars at early times are already chemically evolved systems with metal-enriched BLR (reaching metallicities close to the solar value $Z \sim Z_{\odot}$). In fact, the FeII/MgII ratio of quasars shows little evolution over cosmic time up to very high redshifts ($z \sim 7$), suggesting rapid BLR enrichment at early cosmic epochs \citep{Fan_et_2023, Jiang_et_2024}. This seems to be in stark contrast with lower luminosity JWST-AGN, which apparently have not yet attained chemical maturity and still have low BLR metallicities. Therefore the metallicity may play a major role in determining the AGN properties --such as the strength of radiative dusty feedback-- in the early Universe. 

On the other hand, JWST-AGN share intriguing similarities with metal-poor dwarf galaxies hosting AGN in the local Universe. Common properties observed in both classes include X-ray/radio weakness, Balmer absorption features, and low $R_\mathrm{FeII}$ ratios \citep{Burke_et_2021, Mazzolari_et_2024b, Trefoloni_et_2024}. In particular, remarkable similarities between GN-28074 and SBS0335 (a local metal-poor dwarf galaxy) have been highlighted \citep{Juodzbalis_et_2024}. These unique features could be ultimately driven by the low metallicity characterising both local dwarf galaxies and early JWST-AGN. In this perspective, nearby dwarf galaxies may be considered as local analogs of JWST-AGN in the early Universe, albeit further studies are required to understand their true nature 
and any potential evolutionary link. 


\section*{Acknowledgements}

We thank the anonymous referee for their careful reading and suggestions. 


\section*{Data availability}

No new data were generated or analysed in support of this research.


\bibliographystyle{mn2e}
\bibliography{biblio.bib}

\label{lastpage}

\end{document}